\documentclass[draft,onecolumn,12pt]{IEEEtran}

\usepackage{amsmath}   

 \newtheorem{thm}{Theorem}
 \newtheorem{cor}{Corollary}
 \newtheorem{lem}{Lemma}
 \newtheorem{prop}{Proposition}
 \newtheorem{defn}{Definition}

 \newtheorem{exam}{Example}
 \newenvironment{prf}{{\emph{Proof: }}}
 {\hfill QED.}

\begin{document}
%
\title{New Constructions of Permutation Arrays}
%
%
\author{Lizhen Yang,
        Kefei Chen, Luo Yuan
\thanks{Manuscript received August 29, 2006.
        This work was supported by NSFC under grants 90104005 and 60573030.}
\thanks{Lizhen Yang is with the department of computer science and
engineering, Shanghai Jiaotong University, 800 DongChuan Road,
Shanghai, 200420, R.P. China (fax: 86-021-34204221, email:
lizhen\_yang@msn.com).}
\thanks{Kefei Chen is with the department of computer science and
engineering, Shanghai Jiaotong University, 800 DongChuan Road,
Shanghai, 200420, R.P. China (fax: 86-021-34204221, email:
Chen-kf@sjtu.edu.cn).}}
%
%
%
\markboth{Journal of \LaTeX\ Class Files,~Vol.~1,
No.~11,~August~2006}{Shell \MakeLowercase{\textit{et al.}}: Bare
Demo of IEEEtran.cls for Journals}
%



\maketitle

\begin{abstract}
A permutation array(permutation code, PA) of length $n$ and distance
$d$, denoted by $(n,d)$ PA, is a set of permutations $C$ from some
fixed set of $n$ elements such that the Hamming distance between
distinct members $\mathbf{x},\mathbf{y}\in C$ is at least $d$. In
this correspondence, we present two constructions of PA from
fractional polynomials over finite field, and a construction of
$(n,d)$ PA from permutation group with degree $n$ and minimal degree
$d$. All these new constructions produces some new lower bounds for
PA.
\end{abstract}

%
\IEEEpeerreviewmaketitle

\section{Introduction}
\PARstart{L}{e}t $\Omega$ be an arbitrary nonempty infinite set, and
$Sym(\Omega)$ denote the symmetric group formed by the permutations
over $\Omega$. Two distinct permutations $\mathbf{x},\mathbf{y}\in
Sym(\Omega)$ have distance $d$ if $\mathbf{x}\mathbf{y}^{-1}$ has
exactly $d$ unfixed points, in other words, there are exactly $d$
points $\alpha \in \Omega$ such that $\mathbf{x}(\alpha)\neq
\mathbf{y}(\alpha)$. This distance is also called Hamming distance.
A permutation array(permutation code, PA) of length $n$ and distance
$d$, denoted by $(n,d)$ PA, is a set of permutations $C$ from some
fixed set of $n$ elements such that the distance between distinct
members $\mathbf{x},\mathbf{y}\in C$ is at least $d$. An $(n,d)$ PA
of size $M$ is called an $(n,M,d)$ PA. The maximum size of an
$(n,d)$ PA is denoted as $P(n,d)$.

PAs are somewhat studies in the 1970s. In 2000, an application by
Vinck ~\cite{Ferreira00,Vinck00Code,Vinck00Coded,Vinck00Coding} of
PAs to a coding/modulation scheme for communication over power lines
has created renewed interest in PAs. But the constructions and
bounds for PAs are far from completely. In this correspondence, we
focus on the constructions of PAs. Several papers have been devoted
to this problem. In 1974 Blake~\cite{Blake74} presented a
construction of PA based on the sharply $k-$transitive groups. Using
this method, he constructed $(n,3),(11,8),(12,8),(q,q-1),(1+q,q-1)$
PAs with maximum sizes, where $q$ is the power of a prime. In 2000
Kl$\o$ve~\cite{Klove00combin} gave some constructions of $(n,n-1)$
PA using the linear maps of a ring(commutative with unity). In 2001
Wadayama and Vinck~\cite{Wadayama01} presented some multilevel
constructions of PAs with length $n=2^{m}$. In 2002 Ding, et al.
~\cite{Ding20} presented a method to construct an $r-$bounded
$(mn,mn-uv)$ PA from an $r-$bounded $(n,n-u)$ PA and an
$s$-seperable $(m,m-v)$ PA. In 2003, Chang, et al.~\cite{Chang03}
presented the distance-preserving mapping to construct PAs from
binary codes. In 2004, Fu and Kl{\o}ve ~\cite{Fu-fang-wei04}
presented two constructions of PAs from PAs and $q-$ary codes, Chu,
et al.~\cite{wensong04} gave several constructions, including
construction from permutation polynomials over finite fields,
Colbourn, et al.~\cite{Colbourn04} constructed $(n,n-1)$ PAs from
mutually orthogonal latin squares of order $n$.

In this correspondence, we present two constructions of PA from
fractional polynomials over finite fields, and a construction of
$(n,d)$ PA from permutation group with degree $n$ and minimal
degree $d$. All these new constructions yieds some new lower
bounds for PA.

In the rest of this correspondence, we always denote $q$ as a
power of prime.

\section{Construction of PAs from fractional polynomials}
Polynomials over finite fields are often used to construct codes.
In~\cite{wensong04}, a class of polynomials called permutation
polynomials are directly applied to construction of PAs.  Let
$F_q$ be a finite field of order $q$. A polynomial $f$ over $F_q$
is said to be a permutation polynomial (PP) if the induced map
$\alpha\mapsto f(\alpha)$ from $F_q$ to itself is bijective. Let
$N_k(q)=\{f:f\in F_q[x]\mbox{ is a PP, }\partial(f)=k\}$. It was
shown in ~\cite{wensong04} that the set of all of the permutation
polynomials over $F_q$ with degree $\leq d$ forms an $(n,n-d)$ PA.
\begin{thm}~\cite{wensong04}.
\[
P(q,q-d)\geq \sum_{i=0}^d N_i(q).
\]
\end{thm}

The set of monic permutation polynomials over $F_q$ with degree
$\leq d+1$ forms an $(q,q-d)$ PA also.
\begin{thm}~\cite{wensong04}.
Suppose there are $E$ monic permutation polynomials over $F_q$ of
degree less than or equal to $d+1$. Then $P(q,q-d)\geq E$.
\end{thm}

In this section, we present two constructions of $(q,d)$ and
$(q+1,d)$ based on the fractional polynomials over finite fields.

\subsection{Construction of PAs with length $q$}
\begin{defn}
A fractional polynomial over $F_q$ is of form $\frac{f(x)}{g(x)}$,
where $f(x),g(x)$ are polynomials over $F_q$. Two fractional
polynomials $\frac{f_1(x)}{g_1(x)}$ and $\frac{f_2(x)}{g_2(x)}$ are
said to be equal, denoted as
$\frac{f_1(x)}{g_1(x)}=\frac{f_2(x)}{g_2(x)}$, if and only if
$f_1(x)=f_2(x),g_1(x)=g_2(x)$. If $g(x)$ is monic and
$(f(x),g(x))=1$, then we say $\frac{f(x)}{g(x)}$ is sub-normalized
fractional polynomial. Let $SFP(q)$ denote the set of all
sub-normalized fractional polynomials over $F_q$. For
$\frac{f(x)}{g(x)}\in SFP(q)$, define
\[
V\left(\frac{f(x)}{g(x)}\right)=\left|\left\{\frac{f(\alpha)}{g(\alpha)}:\alpha\in
F_q,g(\alpha)\neq 0\right\}\right|.
\]
\end{defn}

\begin{lem}\label{lem:fractonalpolynomial:nonzero}
Suppose that $\phi=\frac{f_1(x)}{g_1(x)}$ and
$\psi=\frac{f_2(x)}{g_2(x)}$ are sub-normalized fractional
permutations over $F_q$ such that
\[
\partial(f_1(x)g_2(x))\leq q-2, \partial(f_2(x)g_1(x))\leq q-2.
\]
Then
\[
f_1(x)g_2(x)-f_2(x)g_1(x)=0.
\]
if and only if $\phi=\psi$.
\end{lem}
\begin{prf}The sufficiency is clear, we need only to prove the
necessity. If $f_1(x)g_2(x)-f_2(x)g_1(x)=0$, then
$f_1(x)g_2(x)=f_2(x)g_1(x)$, moreover $\partial(f_1(x)g_2(x))\leq
q-2,
\partial(f_2(x)g_1(x))\leq q-2$ and $(f_1(x),g_1(x))=1$, then
$g_1(x)|g_2(x)$. Similarly, $g_2(x)|g_1(x)$. Hence $g_1(x)=g_2(x)$
because $g_1(x)$ and $g_2(x)$ are monic. Then $f_1(x)=f_2(x)$
follows immediately.
\end{prf}

\begin{defn}
A PA-mapping for length $q$ (for short: an $q-$PAM) is a mapping
\begin{eqnarray*}
&\pi:& SFP(q)\mapsto Sym(F_q)\\
&&\frac{f(x)}{g(x)}\mapsto \psi
\end{eqnarray*}
such that for each $\alpha\in F_q$, if
\[
A=\{\beta\in F_q:g(\beta)\neq
0,\frac{f(\beta)}{g(\beta)}=\alpha\}\neq \O,
\]
then $\psi^{-1}(\alpha)\in A$, in other words, there exists
$\beta'\in A$ satisfying $\psi(\beta')=\alpha$.
\end{defn}

\begin{prop}
The number of $q-$PAMs is at least
\[
\prod_{\phi\in SFP(q)}(q-V(\phi))!.
\]
\end{prop}
\begin{prf}
We can construct a $q-$PAM  $\pi$ as follows. For each
$\phi=\frac{f(x)}{g(x)}\in SPF(q)$, according to the definition of
$q-$PAM, we choose $V(\phi)$ members of
$\{\pi(\phi)(\alpha):\alpha\in F_q\}$ determined by
$f(\alpha),g(\alpha)$, namely
$\pi(\phi)(\alpha)=\frac{f(\alpha)}{g(\alpha)}$,  and set the other
$q-V(\phi)$ members of $\{\pi(\phi)(\alpha):\alpha\in F_q\}$ to be
any possibilities satisfying $\{\pi(\phi)(\alpha):\alpha\in
F_q\}=F_q$. There are at least $(q-V(\phi))!$ possibilities of
$\{\pi(\phi)(\alpha):\alpha\in F_q\}$.  Thus we complete the proof.
\end{prf}

\begin{lem}\label{lem:s,t}
\[
\min\{s-s_1,t-t_1\}+\min\{s-s_2,t-t_2\}+\max\{s_1+t_2,s_2+t_1\}\leq
s+t.
\]
\end{lem}
\begin{prf}
For case $\max\{s_1+t_2,s_2+t_1\}=s_1+t_2$, we have
\begin{eqnarray*}
&&\min\{s-s_1,t-t_1\}+\min\{s-s_2,t-t_2\}+\max\{s_1+t_2,s_2+t_1\}\\
&=&\min\{s-s_1,t-t_1\}+\min\{s-s_2,t-t_2\}+(s_1+t_2)\\
&\leq&(s-s_1)+(t-t_2)+(s_1+t_2)\\
&=&s+t.
\end{eqnarray*}
Similarly, for case that $\max\{s_1+t_2,s_2+t_1\}=s_2+t_1$, the
statement holds also.
\end{prf}

\begin{defn}
Let $s,t$ be non-negative integer constants satisfying $s+t\leq
q-2$. Then we define $SFP(q,s,t)$ be the set of all
$\frac{f(x)}{g(x)}\in SFP(q)$ with $\partial(f(x))=s'\leq s$,
$\partial(g(x))=t'\leq t$ and
$$q-V\left(\frac{f(x)}{g(x)}\right)\leq \min\{s-s',t-t'\}.$$
\end{defn}

By definition, $SFP(q,s,0)$ is equivalent to the set of all
permutation polynomials with degree $\leq s$. In this point,
$SFP(q,s,t)$ can be regarded as a generalization of permutation
polynomials, however, $SFP(q,s,t)$ are used to construct PAs with
help of $q-$PAM, rather than directly construction.
\begin{thm}\label{thm:q:SFP}
Let $s,t$ be non-negative integer constants satisfying $s+t\leq
q-2$. Then for any $q-$PAM $\pi$, $\{\pi(\phi):\phi\in
SFP(q,s,t)\}$ is a $(q,|SFP(q,s,t)|,q-s-t)$ PA.
\end{thm}
\begin{prf}
Let $\phi_1=\frac{f_1(x)}{g_1(x)},\phi_2=\frac{f_2(x)}{g_2(x)}\in
SFP(q,s,t)$ with $\phi_1\neq \phi_2$,
$\partial(f_1(x))=s_1,\partial(g_1(x))=t_1,\partial(f_2(x))=s_2,\partial(g_2(x))=t_2$.

Let $r=|\{\alpha\in F_q:g_1(\alpha)\neq 0,g_2(\alpha)\neq
0,\pi(\phi_1)(\alpha)=\frac{f_1(\alpha)}{g_1(\alpha)}=\pi(\phi_2)(\alpha)=\frac{f_2(\alpha)}{g_2(\alpha)}\}|$,
then
\begin{eqnarray*}
r&\leq&\left|\left\{\alpha\in F_q:g_1(\alpha)\neq
0,g_2(\alpha)\neq
0,\frac{f_1(\alpha)}{g_1(\alpha)}=\frac{f_2(\alpha)}{g_2(\alpha)}\right\}\right|\\
&=&\left|\left\{\alpha\in F_q:g_1(\alpha)\neq 0,g_2(\alpha)\neq
0,f_1(\alpha)g_2(\alpha)-f_2(\alpha)g_1(\alpha)=0\right\}\right|,
\end{eqnarray*}
by Lemma~\ref{lem:fractonalpolynomial:nonzero},
$f_1(x)g_2(x)-f_2(x)g_1(x)\neq 0$, then
\begin{eqnarray*}
r&\leq&\partial(f_1(x)g_2(x)-f_2(x)g_1(x))\\
&\leq&\max\{s_1+t_2,s_2+t_1\}.
\end{eqnarray*}

Then by the definition of $q-$PAM, the number of roots of
$\pi(\phi_1)(x)-\pi(\phi_2)(x)=0$ in $F_q$ is at most
\begin{eqnarray*}
&&  (q-V(\phi_1(x)))+(q-V(\phi_2(x)))+r\\
&&\leq
\min\{s-s_1,t-t_1\}+\min\{s-s_2,t-t_2\}+\max\{s_1+t_2,s_2+t_1\}\\
&&\leq s+t,
\end{eqnarray*}
where the last inequality follows from Lemma~\ref{lem:s,t}. This
yields the theorem.
\end{prf}

\begin{cor}For $k\leq q-2$,
$$PA(q,q-k)\geq \max\{|SPF(q,s,t)|:s\geq 0,t\geq 0,s+t=k\}
\geq \sum_{i=0}^{k}N_i(q).$$
\end{cor}

Unfortunately, even the enumeration of permutation polynomials are
far from complete ( $N_k(q)$ has been known for $q\leq 5$
~\cite{Lidl97}), the enumeration of $SFP(q,s,t)$ seem more
difficulty than  that of permutation polynomials. But for small
values of $q,s,t$, we can search by computer by checking all
$\frac{f(x)}{g(x)}\in SFP(q)$ with $\partial(f(x))\leq s$ and
$\partial(g(x))\leq t$. Similar to case of  permutation
polynomials, we can reduce the complexity of checking tasks by
normalized their forms.
\begin{defn}
A fractional polynomial $\frac{f(x)}{g(x)}$ over $F_q$ is said to
be normalized if  $(f(x),g(x))=1$, both $f(x)$ and $g(x)$ are
monic, and when the degree $s$ of $f$ is not divisible by the
characteristic of $F_q$, the coefficient of $x^{s-1}$ is 0.
\end{defn}

Let $\frac{f(x)}{g(x)}\in SFP(q,s,t)$. For $\alpha,\beta\in
F_q,\alpha\neq 0$, then $\psi=\frac{\alpha
f(x+\beta)}{g(x+\beta)}\in SFP(q,s,t)$ is again a member of
$SFP(q,s,t)$. By choosing $\alpha,\beta$ suitably, we can obtain
$\psi$ in normalized form. For a given normalized fractional
polynomial $\frac{f(x)}{g(x)}$, the number of distinct such
$\frac{\alpha f(x+\beta)}{g(x+\beta)}$ is either $q(q-1)$ or
$q-1$, depending on whether $(q,i)=1$ for some $i\geq 1$ such that
there is a nonzero coefficient of $x^i$. By this approach, we find
some new lower bounds for PAs by computer. For $q\leq 23$ be prime
and $k\leq 5$, the new lower bounds on $P(q,q-k)$ are as
following:
\begin{equation*}
\begin{array}{lll}
P(19,16)\geq 684, & P(19,15)\geq 6840, & P(19,14)\geq 65322.
\end{array}
\end{equation*}

\emph{Note: } We also find the two bounds $P(19,16)\geq 684,
P(19,15)\geq 6840$~\cite{Yang-lowerbound06} from another method.

\subsection{Construction of PAs with length $q+1$}
\begin{defn}
A PA-mapping for length $q+1$ (for short: an $(q+1)-$PAM) is a
mapping
\begin{eqnarray*}
&\pi:&SFP(q)\mapsto Sym(F_q\cup\{\infty\})\\
&&\frac{f(x)}{g(x)}\mapsto \psi
\end{eqnarray*}
such that:

(1)for each $\alpha\in F_q$,
\[
A=\{\beta:\beta\in F_q,g(\beta)\neq
0,\frac{f(\beta)}{g(\beta)}=\alpha\}\neq \O,
\]
$\psi^{-1}(\alpha)\in A$, in other words, there exists
 $\beta'\in A$ satisfying $\psi(\beta')=\alpha$;

(2)if $g(x)=0$ has no root in $F_{q}$, then $\psi(\infty)=\infty$,
else $\psi^{-1}(\infty)$ is a root of $g(x)=0$ in $F_q$.
\end{defn}

\begin{prop}
The number of $(q+1)-$PAMs is at least
\[
\prod_{\phi\in SFP(q)}(q-V(\phi))!.
\]
\end{prop}
\begin{prf}
We can construct a $(q+1)-$PAM  $\pi$ as follows. For each
$\phi=\frac{f(x)}{g(x)}\in SPF(q)$, according to the definition of
$(q+1)-$PAM, we choose $V(\phi)$ members of
$\{\pi(\phi)(\alpha):\alpha\in F_q\}$ determined by
$f(\alpha),g(\alpha)$, namely
$\pi(\phi)(\alpha)=\frac{f(\alpha)}{g(\alpha)}$, and choose a
members of $\{\pi(\phi)(\alpha):\alpha\in F_q\cup\{\infty\}\}$ equal
to $\infty$,  and then set the other $q-V(\phi)$ members of
$\{\pi(\phi)(\alpha):\alpha\in F_q\cup\{\infty\}\}$ to be any
possibilities satisfying $\{\pi(\phi)(\alpha):\alpha\in
F_q\cup\{\infty\}\}=F_q\cup\{\infty\}$. There are at least
$(q-V(\phi))!$ possibilities of $\{\pi(\phi)(\alpha):\alpha\in
F_q\cup\{\infty\}\}$. Thus we complete the proof.
\end{prf}

\begin{defn}
Let $s,t,a,b$ be integer constants satisfying
\begin{eqnarray*}
&&s\geq 0,t\geq 0,s+a\geq 0,t+b\geq 0,\\
&&s+t\leq q-2,\\
&&s+t+a\leq q-2,\\
&&s+t+b\leq q-2,\\
&&s+t+a+b\leq q-2.\\
\end{eqnarray*}
Then we define $SFP(q,s,t,a,b)$ be set of all $\frac{f(x)}{g(x)}\in
SFP(q)$ such that, supposing $s'=\partial(f(x))$,
$t'=\partial(g(x)),v=V\left(\frac{f(x)}{g(x)}\right)$, for case that
$g(x)=0$ has roots in $F_q$,
\[
s'\leq s,t'\leq t,
\]
and
\[
q-v\leq \min\{s-s',t-t'\}+1;
\]
for case that $g(x)=0$ has no root in $F_q$,
\[
s'\leq s+a,t'\leq t+b
\]
and
\[
q-v\leq \min\{s+a-s',t+b-t'\}.
\]
\end{defn}

\begin{thm}\label{thm:SFP:q+1}
Let $s,t,a,b,d$ be integer constants satisfying
\begin{eqnarray*}
&&s\geq 0,t\geq 0,s+a\geq 0,t+b\geq 0,\\
&&s+t\leq q-2,\\
&&s+t+a\leq q-2,\\
&&s+t+b\leq q-2,\\
&&s+t+a+b\leq q-2,\\
&&d=\min\{q-s-t,q-s-t-a-b,q+1-s-t-\max\{a,b\}\}.
\end{eqnarray*}
Then for any $(q+1)-$PAM $\pi$, $\{\pi(\phi):\phi\in
SFP(q,s,t,a,b)\}$ is a $(q+1,d)$ PA with size $|SFP(q,s,t,a,b)|$.
\end{thm}
\begin{prf}
Let $\phi_1=\frac{f_1(x)}{g_1(x)},\phi_2=\frac{f_2(x)}{g_2(x)}\in
 SFP(q,s,t,a,b)$ with $\phi_1\neq \phi_2$,
$s_1=\partial(f_1(x)),t_1=\partial(g_1(x)),v_1=V(\phi_1)$,
$s_2=\partial(f_2(x)),t_2=\partial(g_2(x)),v_2=V(\phi_2)$.

Let $r=|\{\alpha\in F_q:g_1(\alpha)\neq 0,g_2(\alpha)\neq
0,\pi(\phi_1)(\alpha)=
\frac{f_1(\alpha)}{g_1(\alpha)}=\pi(\phi_2)(\alpha)=\frac{f_2(\alpha)}{g_2(\alpha)}\}|$,
then
\begin{eqnarray*}
r&\leq&\left|\left\{\alpha\in F_q:g_1(\alpha)\neq
0,g_2(\alpha)\neq
0,\frac{f_1(\alpha)}{g_1(\alpha)}=\frac{f_2(\alpha)}{g_2(\alpha)}\right\}\right|\\
&=&\left|\left\{\alpha\in F_q:g_1(\alpha)\neq 0,g_2(\alpha)\neq
0,f_1(\alpha)g_2(\alpha)-f_2(\alpha)g_1(\alpha)=0\right\}\right|
\end{eqnarray*}
By Lemma~\ref{lem:fractonalpolynomial:nonzero},
$f_1(x)g_2(x)-f_2(x)g_1(x)\neq 0$, then
\begin{eqnarray*}
r&\leq&\partial(f_1(x)g_2(x)-f_2(x)g_1(x))-z\\
&\leq&\max\{s_1+t_2,s_2+t_1\}-z,
\end{eqnarray*}
where $z=\{\alpha\in F_q:g_1(\alpha)=g_2(\alpha)=0\}$. Now we are
ready to find the upper bound  on the number of roots of
$\pi(\phi_1)(x)-\pi(\phi_2)(x)=0$ in $F_q\cup\{\infty\}$, which is
denoted as $R$. We discuss in four cases:

Case I): Both $g_1(x)=0$ and $g_2(x)=0$ have roots in $F_q$. We
further discuss in two subcases:

Subcase 1):$g_1(x)=0$ and $g_2(x)=0$ have at least a common root in
$F_q$, namely $z\geq 1$. By the definition of $(q+1)-$PAM, we have
\begin{eqnarray*}
R&\leq& z+(q-v_1-z)+(q-v_2-z)+r+1\\
&\leq&z+(q-v_1-z)+(q-v_2-z)+\max\{s_1+t_2,s_2+t_1\}-z+1\\
&=&(q-v_1)+(q-v_2)+\max\{s_1+t_2,s_2+t_1\}-2z+1\\
&\leq&\min\{s-s_1,t-t_1\}+1+\min\{s-s_2,t-t_2\}+1+\max\{s_1+t_2,s_2+t_1\}-1\\
&\leq&s+t+1.
\end{eqnarray*}
where the last inequality follows from Lemma~\ref{lem:s,t}.

Subcase 2):$z=0$. Then if $\alpha\in Z_q$ is a root of $g_1(x)=0$
satisfying $\pi(\phi_1)(\alpha)=\infty$ then
$\pi(\phi_2)(\alpha)\neq \infty$, whereas if $\alpha'\in Z_q$ is a
root of $g_2(x)=0$ satisfying $\pi(\phi_2)(\alpha')=\infty$ then
$\pi(\phi_1)(\alpha')\neq \infty$. So we have
\begin{eqnarray*}
  R &\leq& (q-v_1-1)+(q-v_2-1)+r+1\\
&\leq&\min\{s-s_1,t-t_1\}+\min\{s-s_2,t-t_2\}+\max\{s_1+t_2,s_2+t_1\}+1\\
&\leq&s+t+1.
\end{eqnarray*}

Case II): Both $g_1(x)=0$ and $g_2(x)=0$ have no root, then $z=0$.
By the definition of $(q+1)-$PAM, we have
\begin{eqnarray*}
R&\leq&(q-v_1)+(q-v_2)+r+1\\
&\leq&\min\{s+a-s_1,t+b-t_1\}+\min\{s+a-s_2,t+b-t_2\}+\max\{s_1+t_2,s_2+t_1\}+1\\
&\leq&s+a+t+b+1.
\end{eqnarray*}

Case III): $g_1(x)=0$ has roots in $F_q$ while $g_2(x)=0$ has no
root. Then $z=0$. If $\alpha\in Z_q$ is a root of $g_1(x)=0$
satisfying $\pi(\phi_1)(\alpha)=\infty$ then
$\pi(\phi_2)(\alpha)\neq \infty$, and $\pi(\phi_1)(\infty)\neq
\infty$ while $\pi(\phi_2)(\infty)= \infty$. Then by the
definition of $(q+1)-$PAM, we have
\begin{eqnarray*}
R&\leq&(q-v_1-1)+(q-v_2)+r\\
&\leq&\min\{s-s_1,t-t_1\}+\min\{s+a-s_2,t+b-t_2\}+\max\{s_1+t_2,s_2+t_1\}\\
&\leq&\min\{s-s_1,t-t_1\}+\min\{s-(s_2-a),t-(t_2-b)\}\\
&&\qquad+\max\{s_1+(t_2-b),(s_2-a)+t_1\}+\max\{a,b\}\\
&\leq&s+t+\max\{a,b\}.
\end{eqnarray*}

Case IV): $g_1(x)=0$ has no roots while $g_2(x)=0$ has roots in
$F_q$. It can be proved that $R\leq s+t+\max\{a,b\}$ similar to
Case III.

Now we can conclude that $\{\pi(\phi):\phi\in SFP(q,s,t,a,b)\}$ is a
$(q+1,d)$ PA with size $|SFP(q,s,t,a,b)|$, where
\[
d=q+1-R\geq \min\{q-s-t,q-s-t-a-b,q+1-s-t-\max\{a,b\}\}.
\]
\end{prf}

Comparing the definitions of $SFP(q,s,t)$ with $SFP(q,s,t,a,b)$, we
find $SFP(q,s,t)\subseteq SFP(q,s,t,0,0)$. This in conjunction with
Theorem~\ref{thm:SFP:q+1} implies the following Corollary.

\begin{cor}For $k+1\leq q-2$,\small
\begin{eqnarray*}
&&PA(q+1,q-k)\geq \\
&&\max\{|SFP(q,s,t,a,b)|:s+t=k,s\geq 0,t\geq 0,s+a\geq 0,t+b\geq
0,(a,b)\in\{(0,0),(1,-1),(-1,1)\}\}\\
&&\geq \max\{|SPF(q,s,t):s\geq 0,t\geq 0,s+t=k\}.
\end{eqnarray*}
\end{cor}

As the case of $SFP(q,s,t)$, the enumeration of $SFP(q,s,t,a,b)$
is difficulty to determined, while for small values of $q,s,t$, we
can find by computer by checking all $\frac{f(x)}{g(x)}\in SFP(q)$
with $\partial(f(x))\leq s+\max\{0,a\},\partial(f(x))\leq
t+\max\{0,b\}$. The complexity of checking task can be also
reduced by only checking the normalized forms. Let
$\frac{f(x)}{g(x)}\in SFP(q,s,t,a,b)$. For $\alpha,\beta\in
F_q,\alpha\neq 0$, then $\psi=\frac{\alpha
f(x+\beta)}{g(x+\beta)}\in SFP(q,s,t,a,b)$ is again a member of
$SFP(q,s,t,a,b)$. By choosing $\alpha,\beta$ suitably, we can
obtain $\psi$ in normalized form. For a given normalized
fractional polynomial $\frac{f(x)}{g(x)}$, the number of distinct
such $\frac{\alpha f(x+\beta)}{g(x+\beta)}$ is either $q(q-1)$ or
$q-1$, depending on whether $(q,i)=1$ for some $i\geq 1$ such that
there is a nonzero coefficient of $x^i$. By this approach, we find
some new lower bounds for PA by computer. For $q\leq 23$ be prime
and $k\leq 5$, the new lower bounds on $P(q+1,q-k)$ are as
following:
\begin{equation*}
\begin{array}{lll}
 P(18,14)\geq 9520, & P(20,14)\geq 123804,& P(24,20)\geq 23782.\\
\end{array}
\end{equation*}

\section{Construction of $(n,d)$ PAs from permutation groups with
degree $n$ and minimal degree $d$}

Let $G$ be a finite permutation group with a action on a set
$\Omega$. The order of $G$ is defined as the cardinality of $G$ and
the degree of $G$ is defined as the cardinality of $\Omega$. If
$g\in G$, then the degree of $g$ on $\Omega$ is the number of points
moved by $g$. The minimal degree of $G$ is the minimum degree of a
nontrivial element in $G$. $G$ has fixity $f$ if nontrivial elements
of $G$ fixes $\leq f$ points, and there is a nontrivial element of
$G$ fixing exactly $f$ points. Thus the minimal degree of a
permutation group of degree $n$ and fixity $f$ is $n-f$. By
definitions, for $g_1,g_2\in G$, the distance between $g_1$ and
$g_2$ is the degree of $g_1g^{-1}_2$, this yields the following
theorem immediately.

\begin{thm}~\label{thm:construct:permutation Group}
Let $G$ be a permutation group with degree $n$ and minimal degree
$\geq d$. Then $G$ form an $(n,d)$ PA.
\end{thm}

Then from the knowledge of permutation groups, we can obtain a lot
of PAs  for given lengths and distances.
\begin{exam}
Frobenius groups have fixity one, then Frobenius group of degree
$n$ forms an $(n,n-1)$ PA~\cite[p.85]{Dixon96}. Zassenhaus groups
have fixity two, then Zassenhaus group with degree $n$ forms an
$(n,n-2)$ PAs. The minimal degree of a proper primitive
permutation groups of degree $n$ is at least
$2(\sqrt{n}-1)$~\cite{Liebeck91}, then a proper primitive
permutation group of degree $n$ forms an $(n,2(\sqrt{n}-1))$ PA.
\end{exam}
\begin{exam}
Let $m\geq 5$, $n=m(m-1)/2$, and $d=2m-4$, and permutation group
$G$ be the action of $S_m$ on the set of $2-$sets of
$\{1,2,\ldots,m\}$. This action is primitive of degree
$n=m(m-1)/2$ with minimal degree $d=2m-4$~\cite[Exercise 3.3.5,
p.77]{Dixon96}. Then $G$ is an $(n,d)$ PA with $|G|\geq
\exp(\sqrt{2n}\log\sqrt{2n}-\sqrt{2n})$~\cite[Exercise 5.3.4,
p.155]{Dixon96}.
\end{exam}
\begin{exam}
Let $G$ be the affine group $AGL_d(q)$ acts as a permutation on
the affine space of dimension $d$ over a field of $q$ elements.
Then $G$ is an $(q^d,q^d-q^{d-1})$ PA~\cite[Example 5.4.1,
p.158]{Dixon96} of size $q^{d(d+1)/2}(q^d-1)(q^{d-1}-1)\ldots
(q-1)$. Particularly, for $q=2$, $G$ is a $(2^d,2^{d-1})$ PA with
size $2^{d(d+1)/2}(2^d-1)(2^{d-1}-1)\ldots (2-1)$ which is
$2^{(d+1)/2}$ times the size $2^d(2^d-1)(2^{d-1}-1)\ldots (2-1)$
of PA constructed in ~\cite{Wadayama01}.

\end{exam}

Some new lower bounds for PAs are obtained below.

\begin{lem}
$P(24,16)\geq 244823040,P(23,16)\geq 10200960,P(22,16)\geq
443520$.
\end{lem}
\begin{prf}
 The fixity of Mathieu group $M_{24}$ is at most
$8$~\cite[p.310]{Huppert82}, then $M_{24}$ is a $(24,16)$ PA of
size $|M_{24}|=2^{10}\cdot 3^3\cdot 5\cdot 7\cdot 11\cdot
23=244823040$~\cite[Table 6.1., P.204]{Dixon96}. Since Mathieu
group $M_{23}$ is a one-point stabilizer of $M_{24}$, then
$M_{23}$ is a $(23,16)$ PA with size $|M_{23}|=2^{7}\cdot 3^2\cdot
5\cdot 7\cdot 11\cdot 23=10200960$~\cite[Table 6.1.,
P.204]{Dixon96}. Since Mathieu group $M_{22}$ is a two-point
stabilizer of $M_{24}$, then $M_{22}$ is a $(22,16)$ PA with size
$|M_{22}|=2^{7}\cdot 3^2\cdot 5\cdot 7\cdot 11=443520$~\cite[Table
6.1., P.204]{Dixon96}.
\end{prf}

The permutation group $G$ acting on a set $\Omega$ with
$|\Omega|=n$ is said to be sharply $k$-transitive if, for any two
ordered $k$ subsets of $\Omega$, say $\{i_1,i_2,\ldots,i_k\}$ and
$\{j_1,j_2,\ldots,j_k\}$ there exists exactly one element
$\sigma\in G$ such that $\sigma(i_l)=j_l,l=1,\ldots,k$. The
sharply $k$-transitive group of degree $n$ has been proved in
~\cite{Blake74} that it is an $(n,n-k+1)$ PA of size $n!/(n-k)!$.
 Indeed, a permutation group $G$ with $|G|=n!/(n-k)!$ of degree $n$
is an $(n,n-k+1)$ PA if and only if it is sharply $k$-transitive.

\begin{thm}\label{thm:group:sharply}
A permutation group $G$ with order $n!/(n-k)!$ and degree $n$ is
an $(n,n-k+1)$ PA if and only if it is sharply $k$-transitive.
\end{thm}

\begin{prf}
We need only to prove the necessary. Suppose that $G$ acts on
$\Omega$. For any two ordered $k$ subsets of $\Omega$, say
$\{i_1,i_2,\ldots,i_k\}$, $\{j_1,j_2,\ldots,j_k\}$, there exists
at most one $\sigma\in G$ such that
\begin{equation}\label{eq:sharply}\sigma(i_l)=j_l,l=1,\ldots,k,\end{equation}
since two distinct such permutations would have distance $\leq n-k$.
There are $n!/(n-k)!$ ordered $k$ subsets of $\Omega$, this means
there exists at least one element $\sigma\in G$ satisfying condition
(\ref{eq:sharply}). Hence $G$ is sharply $k$-transitive.
\end{prf}


\hfill mds

\hfill November 18, 2002

\bibliographystyle{IEEEtran}
\bibliography{IEEEabrv,bib}
\end{document}